\documentclass[aps,prd,twocolumn,showpacs,superscriptaddress,groupedaddress]{revtex4}  

\usepackage{amsmath}
\usepackage{amsfonts}
\usepackage{bm}
\usepackage{color}
\usepackage{graphicx}
\definecolor{darkgreen}{cmyk}{0.85,0.2,1.00,0.2}
\usepackage{amsbsy}

\input epsf
\usepackage{amsmath,amssymb,subfigure}
\usepackage{graphicx}
\usepackage{epsfig}
\usepackage{color}
\usepackage{ulem}
\usepackage{epstopdf}
\newcommand{\be}{\begin{eqnarray}}

\newcommand{\ee}{\end{eqnarray}}

\newcommand{\apjl}{Astrophys. J. Lett.}
\newcommand{\aap}{Astron. Astrophys.}
\newcommand{\apjs}{Astrophys. J. Suppl. Ser.}

\newcommand{\ls}{\mathrel{\raise0.27ex\hbox{$<$}\kern-0.70em \lower0.71ex\hbox{{$\scriptstyle \sim$}}}}

\begin{document}

\title{Searching for Oscillations in the Primordial Power Spectrum:  Constraints from Planck (Paper II)}

\author{P.~Daniel Meerburg} 
\affiliation{Department of Astrophysical Sciences, Princeton University, Princeton, NJ 08540 USA. }
\email{meerburg@princeton.edu}
\author{David N. Spergel}
\email{dns@astro.princeton.edu}
\affiliation{Department of Astrophysical Sciences, Princeton University, Princeton, NJ 08540 USA. }
\author{ Benjamin D.~Wandelt}
\email{wandelt@iap.fr}
\affiliation{CNRS-UPMC Univ. Paris 06, UMR7095, Institut d`Astrophysique de Paris, 98bis Bd. Arago, F-75014, Paris, France}
\affiliation{Departments of Physics and Astronomy, University of Illinois at Urbana-Champaign, Urbana, IL 61801, USA}
\date{\today}

\begin{abstract}
In this second of two papers we apply our recently developed code to search for resonance features in the Planck CMB temperature data. We  search both for log spaced oscillations or  linear spaced oscillations and compare
our findings with results of our WMAP9 analysis and the Planck team analysis \citep{2013arXiv1303.5082P}. While there are hints of log spaced resonant features present in the WMAP9 data, the significance of these features weaken with more data.  With more accurate small scale measurements, we also find that  the best fit frequency has shifted and the amplitude has been reduced.  We confirm the presence of a several low frequency peaks, earlier identified by the Planck team, but with a better improvement of fit ($\Delta \chi^2_{\mathrm{eff}} \sim 12$). We further investigate this improvement by allowing the lensing potential to vary as well, showing mild correlation between the amplitude of the oscillations and the lensing amplitude. We find that the improvement of the fit increases even more ($\Delta \chi^2_{\mathrm{eff}} \sim 14$) for the low frequencies that modify the spectrum in a way that mimics the lensing effect. Since these features were not present in the WMAP data, they are primarily due to better measurements of Planck at small angular scales. For linear spaced oscillations we find a maximum $\Delta \chi^2_{\mathrm{eff}} \sim 13$ scanning two orders of magnitude in frequency space, and the biggest improvements are at extremely high frequencies. Again, we recover a best fit frequency very close to the one found in WMAP9, which confirms that the fit improvement is driven by low $\ell$. Further comparisons with WMAP9 show Planck contains many more features, both for linear and log space oscillations, but with a smaller improvement of fit. We discuss the improvement as a function of the number of modes and study the effect of the 217 GHz map, which appears to drive most of the improvement for log spaced oscillations.  Two points strongly suggest that the detected features are fitting a combination of the noise and the dip at $\ell\sim1800$  in the 217 GHz map:  the fit improvement mostly comes from a small range of $\ell$, and  comparison with simulations shows that the fit improvement is consistent with a statistical fluctuation.  We conclude that none of the detected features are statistically significant.
 \end{abstract}

\maketitle
\section{Introduction} 

In this short paper, we will apply our recent introduced method in Ref.~\citep{2013Meerburga} to search for resonant features in the recently released  Planck CMB data. 
We consider two distinct theoretically motivated models:

\be
_1\Delta^2_{\mathcal{R}}(k)  = A_1 \left(\frac{k}{k_*} \right)^{m}\left(1+A_2 \cos [\omega_1 \log k/k_* +\phi_1]\right)\\
\label{eq:powerspectra1}
_2\Delta^2_{\mathcal{R}}(k) = B_1\left(\frac{k}{k_*} \right)^{m}\left(1+B_2 k^n \cos [\omega_2 k +\phi_2]\right)
\label{eq:powerspectra2}
\ee
We refer to the first model as the ``log-spaced oscillations model" and the second model as the ``linear oscillations model".
For example,  axion-monodromy inflation
produces features that can be described by
the logarithmic oscillations model with  $A_1 = H^2/(8\pi^2\epsilon)$, $m=n_s-1$, $A_2 = \delta n_s$, $\omega_1 = -(\phi_*)^{-1}$ and $\phi_1 = \phi_*$. Model that include the effects from a possible boundary on effective field theory (BEFT) predict features that can be described by the linear oscillations model with $B_1 = H^2/(8\pi^2\epsilon)$, $m=n_s-1$, $B_2  = \beta/a_0M$, $n=1$, $\omega_2= 2/a_0 H$ and $\phi_2=\pi/2$. Both initial state modifications and multiverse models \citep{2013JCAP...03..004D} can also produce logarithmic oscillations, while sharp features generate a power spectrum  with linear oscillations (although the amplitude is typically damped as a function of scale). Constraints on oscillations in the WMAP CMB data have been attempted in e.g. Refs.~\citep{2004PhRvD..69h3515M,2011PhRvD..84f3515D, 2007PhRvD..76b3503H, 2009PhRvD..79h3503P, 2012MNRAS.421..369M, 2013arXiv1303.2616P, 2013PhRvD..87h3526A}. Note that model \eqref{eq:powerspectra1} has a unit less frequency while model \eqref{eq:powerspectra2} has units of Mpc. We will omit these units in the rest of the paper for brevity. 

This paper is organized as follows. We present our results on the Planck Data in  \S\ref{analysis} for log and linear spaced oscillations. In  \S\ref{WMAPvsPlanck}, we compare our results with the WMAP9 analysis. We discuss our findings and conclude in  \S\ref{conclusion}.  

\section{Planck Analysis} \label{analysis}

In this analysis, we use a modified version of the publicly available Planck likelihood code \cite{2013arXiv1303.5075P} to search for oscillations in the primordial power spectrum. For this analysis, we found the best fit values for both resonance model parameters and cosmological parameters. We vary all six $\Lambda$CDM parameters plus the phase and the amplitude of the oscillatory correction to the primordial power spectrum while fixing the foreground parameters to their best fit values for the no oscillations model.  The best-fit is found using the Metropolos-Hastings algorithm, which is not the ideal method to look for the best fit, but it does allow us to compute marginalized likelihoods of the parameters and look for potential correlations. 

\vspace{.3 in}
  
\subsection{Log-spaced oscillations}

\begin{figure}[htbp] 
   \centering
   \includegraphics[width=3.4in]{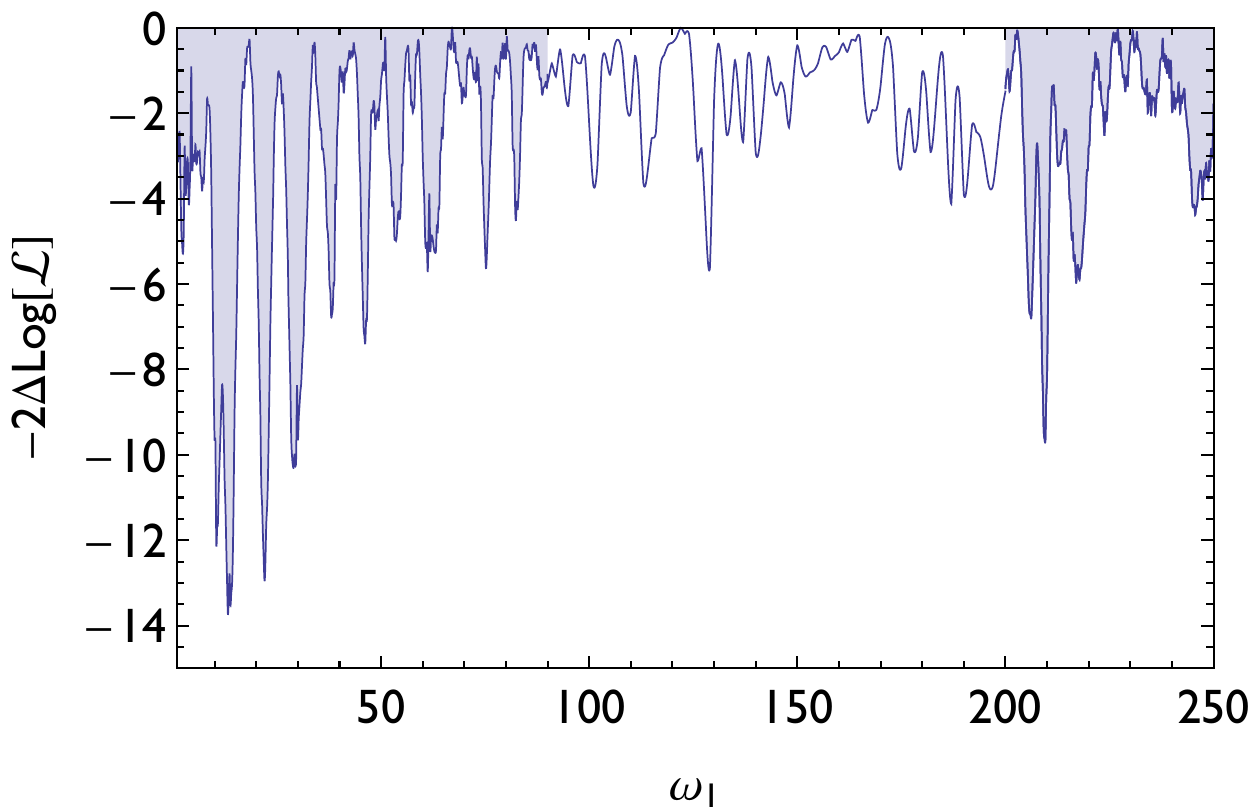} 
   \caption{We plot the improvement of fit versus frequency $\omega_1$ Planck 1 and log spaced oscillations. The biggest improvement are found at low frequencies. Here we allowed $A_{\mathrm{lens}}$ to vary freely. We zoom in on the shaded regions in Fig.~\ref{fig:wmapvsplanck1} and Fig.~\ref{fig:wmapvsplanck2}. }
   \label{fig:planck1}
\end{figure}

\begin{figure}[htbp] 
   \centering
   \includegraphics[width=3.4in]{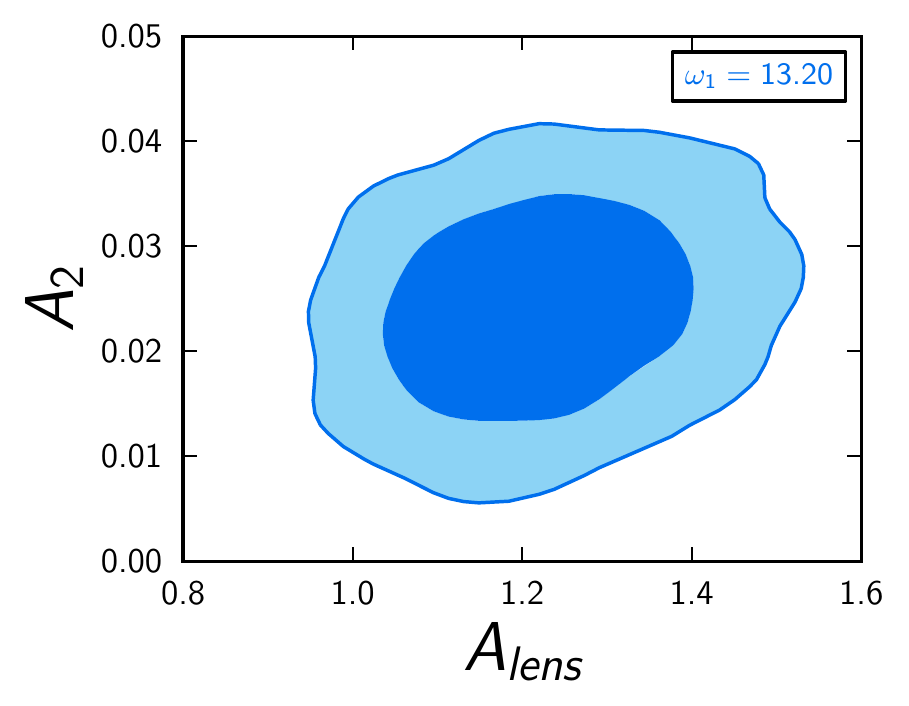} 
   \caption{Marginalized probability distribution of the lensing amplitude versus the amplitude of the (log spaced) oscillations. }
   \label{fig:alens_vs_amp}
\end{figure}

In Fig.~\ref{fig:planck1} shows the improvement in fit as a function of frequency, where the  frequency of the oscillation was varied in 1201 steps between $1 \leq \omega_1 \leq 250$ \footnote{We ran our analysis all the way up to $\omega_1 =300$, with a lower resolution in frequency space, but found no further improvement.}. We observe several frequencies that could be hints of primordial oscillations. We confirm a number of features first observed in Ref.~ \citep{2013arXiv1303.5082P}.  
Our method improves the best fit peak identified by the planck team \citep{2013arXiv1303.5082P} at low frequencies with  $\Delta \chi^2_{\mathrm{eff}} \sim 3$ (with best fit frequency $\omega_1=29.2$) \footnote{After this paper was submitted but before publication, another paper appeared on the archive searching for features \cite{2013PhRvD..88h7302B}. } . After inspecting the resulting fit, we expected some correlation with smooth parameters. We found that varying $A_{\mathrm{lens}}$ enables a further improvement in the fit by another $\Delta \chi^2_{\mathrm{eff}} \sim 2$\footnote{This improvement is with respect to no oscillations, but with lensing turned on.}, but the best fit has shifted towards a lower frequency $\omega_1=13.2$.  In Fig.~\ref{fig:alens_vs_amp} we show the marginalized contour between the lensing amplitude $A_{\mathrm{lens}}$ and the amplitude of the oscillations at the best fit frequency. Further investigation shows that this mild correlation actually shifts slope from peak to peak, which can be explained by the fact that the improvement of fit is at $\ell>1500$ (see  \S\ref{WMAPvsPlanck}); for these low frequencies the contribution to the power spectrum is rather smooth and the lensing amplitude effectively smooths the peak structure. Oscillations can help enhance or reduce this effect, and the correlation coefficient can therefore change signs depending on the phase of the oscillation.

\subsection{Linear-spaced oscillations}
\begin{figure}[htbp] 
   \centering
   \includegraphics[width=3.4in]{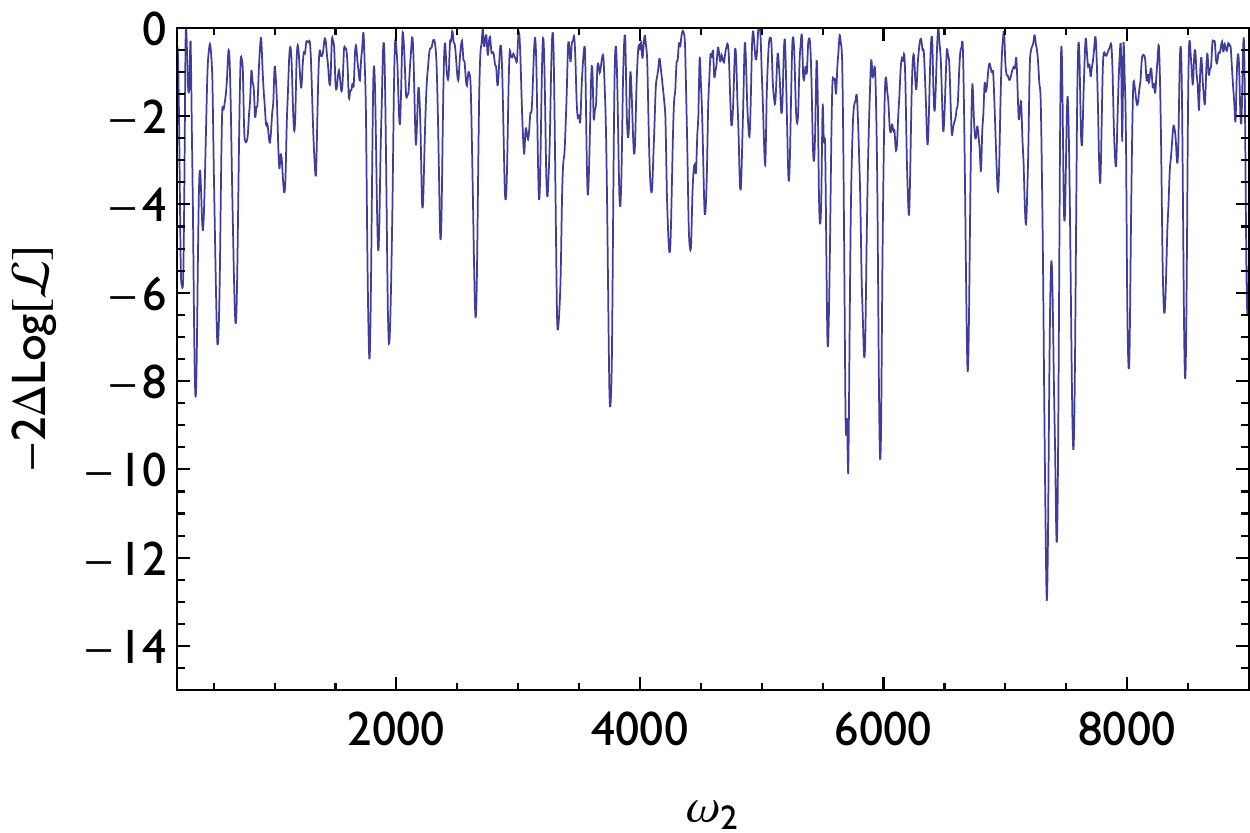} 
   \caption{We plot the improvement of fit versus frequency $\omega_2$ Planck 1 and linear spaced oscillations. }
   \label{fig:planck2}
\end{figure}

In Fig.~\ref{fig:planck2}, we show the improvement as a function of  frequency for
the linear spaced oscillations model.  Again, we vary the phase and amplitude of the oscillation, together with the cosmological parameters, for each of 881 steps in frequency space. For linear spaced oscillations, the resulting improvement is extremely irregular, with no particularly preferred region. The best fit is at a frequency of $\omega_2=7340$, where the $\Delta \chi^2 \sim 13$.

\section{WMAP9 vs Planck1}\label{WMAPvsPlanck}
\subsection{Log-spaced oscillations}
\begin{figure}[htbp] 
   \centering
   \includegraphics[width=3.4in]{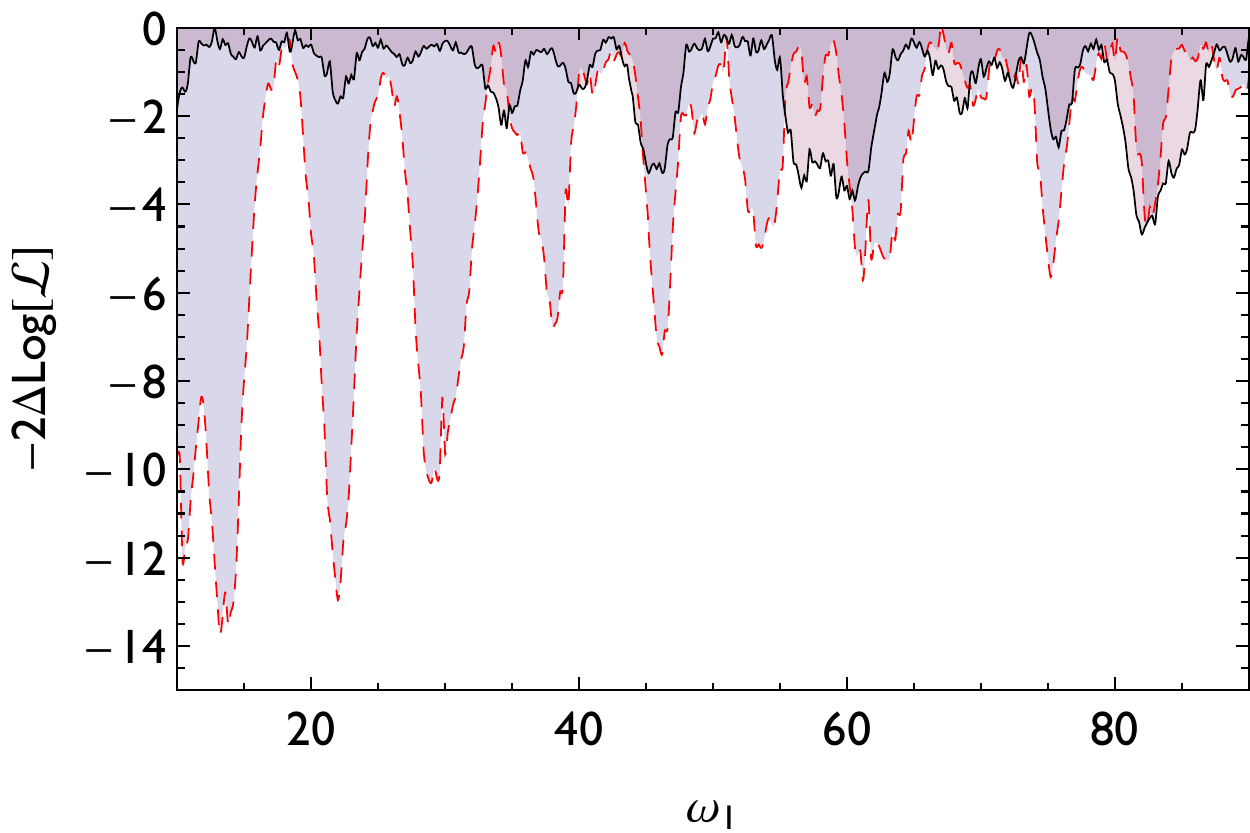} 
   \caption{We plot the improvement of fit for the lowest frequencies for both WMAP9 (solid, black) and Planck 1 (red, dashed). Although some peaks are seen in WMAP 9, Planck has relatively large improvements at the low frequency end, suggesting these are due to the a better fit at small angular scales. We confirm that our code reproduces the findings by the  Planck team, but typically with a bigger improvement in $\chi^2$ due  to allowing both the oscillation parameters and the cosmological parameters to vary.}
   \label{fig:wmapvsplanck1}
\end{figure}

\begin{figure}[htbp] 
   \centering
   \includegraphics[width=3.4in]{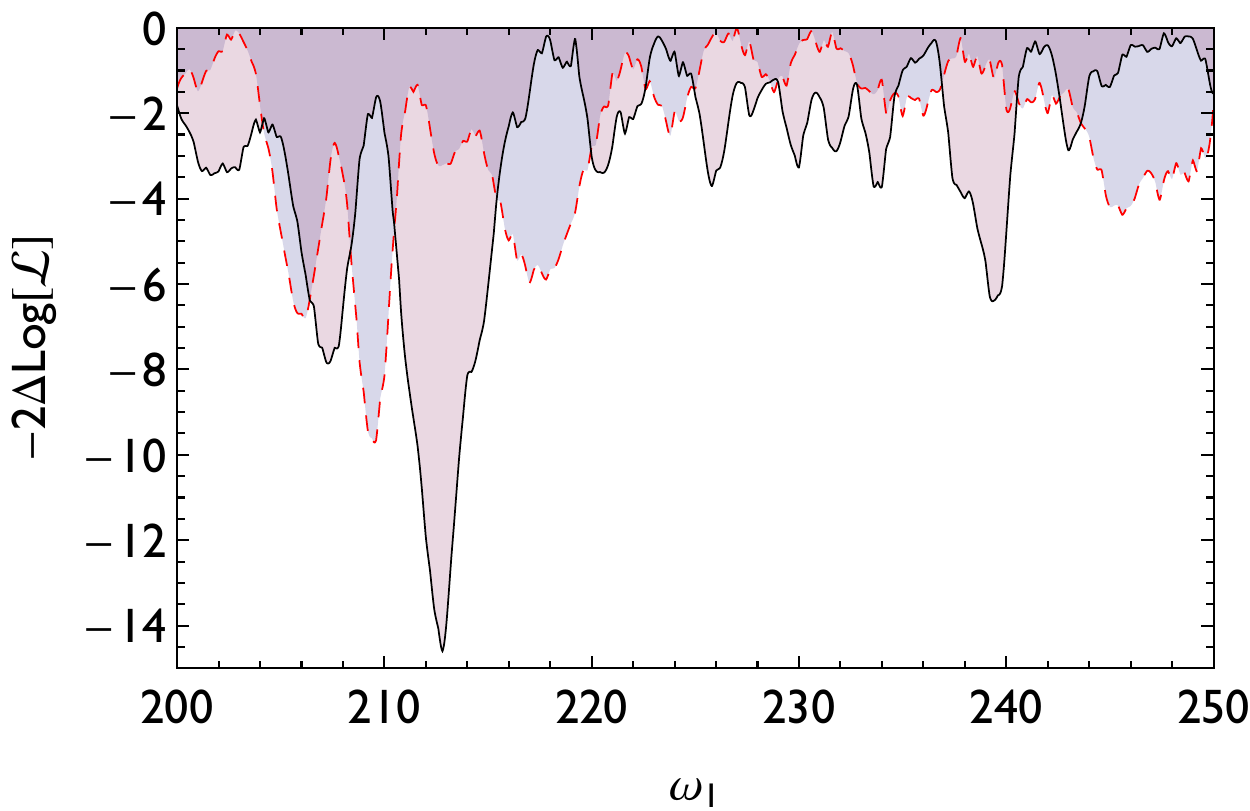} 
   \caption{We plot the improvement of fit for the highest frequencies for both WMAP9 (solid, black) and Planck 1 (red, dashed). }
   \label{fig:wmapvsplanck2}
\end{figure}

In Fig.~\ref{fig:wmapvsplanck1} and Fig.~\ref{fig:wmapvsplanck2} we compare the improvement of fit in our analysis of WMAP data to the improvement in fit
in our analysis of Planck data.  Fig.~\ref{fig:wmapvsplanck1} clearly shows new peaks at the low frequency end in  the Planck data, features that are absent in the WMAP data. At high frequencies, the feature that was seen in the  WMAP analysis is less significant and has shifted. Note that some of this shift is due to different precomputed transfer function, which for high frequencies have a small but non-negligible effect on the projected frequency \citep{2013Meerburga}. For $\ell_{\mathrm{max}} \sim 500$, where WMAP and Planck are both cosmic variance limited, we expect the features to coincide. As
shown in an analysis in Ref.~\cite{2013Flaugera} , the location of the feature shifts as one lowers $\ell_{\mathrm{max}}$ to $ \sim 500$, confirming that a better fit at small scales or high $\ell$ is the source of the shift. 

\begin{figure}[htbp] 
   \centering
   \includegraphics[width=3.4in]{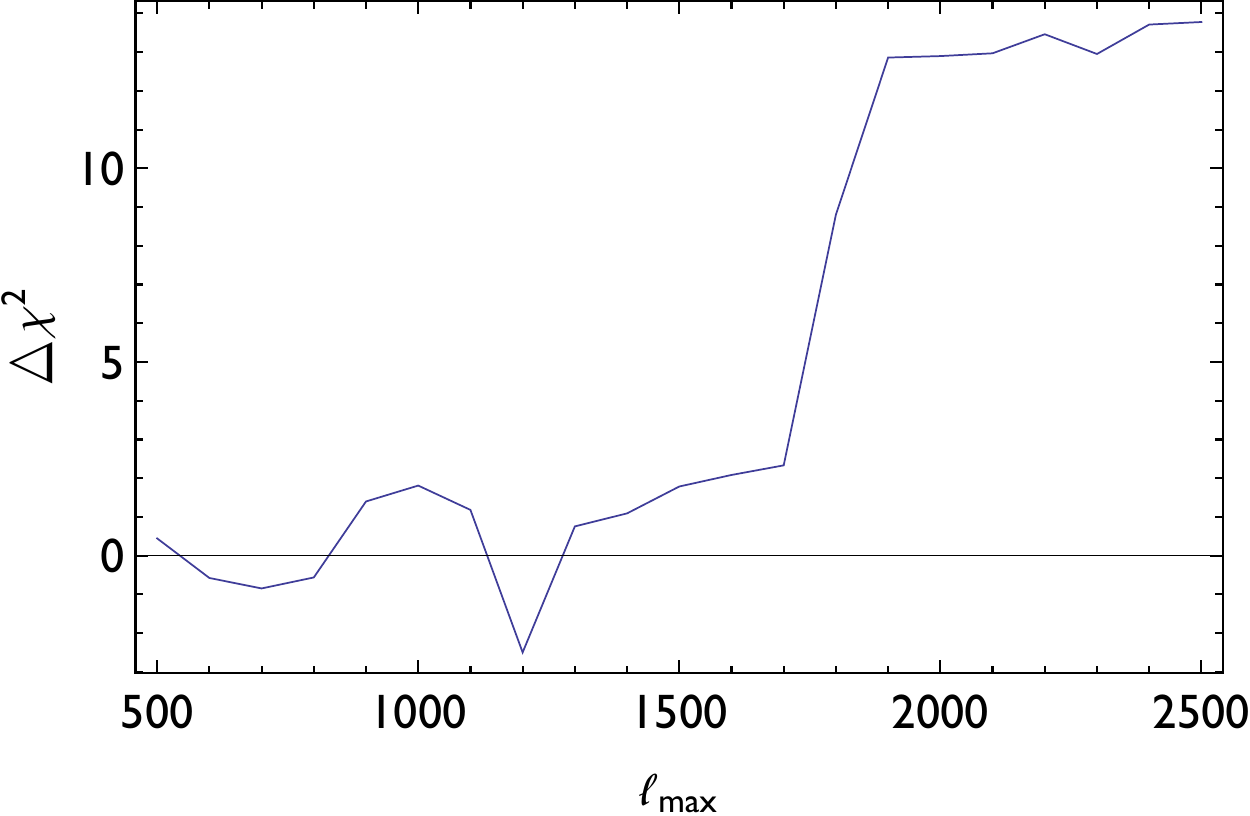} 
   \caption{This figure shows the improvement of the fit versus a reference fit for log spaced oscillations. Two features stand out at  $\ell_{\mathrm{max}}=1800$ and  $\ell_{\mathrm{max}}=1100$. The first of this features is probably caused by the 217 GHz map. Removing this map from the data indeed reduces the improvement to a level comparable to setting $\ell_{\mathrm{max}}=1800$.}
   \label{fig:chivslmax}
\end{figure}

Since the low frequency features are absent in the WMAP data, their  presence should be primarily due to a better fit in the range $500\leq \ell_{\mathrm{max}} \leq 2500$. The correlation between $A_{\mathrm{lens}}$ and the amplitude of the oscillations is a result of the fit being driven by the high $\ell$ Planck data. In Fig. \ref{fig:chivslmax} we show improvement of fit compared to no oscillations as a function of $\ell_{\mathrm{max}}$ for our best fit frequency $\omega_1 = 13.2$. Indeed this plot shows that the improvement comes from multipoles around $\ell=1800$ and $\ell=1100$ (roughly between the 3rd and 4th peak). In the likelihood for our parameters we use the 100 GHz data up to $\ell_{\mathrm{max}} = 1200$, the 143 GHz  data to $\ell_{\mathrm{max}}=2000$ and the 217 GHz data up to $\ell_{\mathrm{max}}=2500$.  As was shown in Ref.~\citep{2013arXiv1303.5076P} the 217 GHz map drives some of the standard $\Lambda$CDM parameters away from their best fit WMAP values and the 217$\times$217 GHz power
spectrum contains a feature at $\ell =1800$ that is not seen in the 143$\times$217 GHz power spectrum. C. By removing the 217 GHz data we find that the improvement drops to $\Delta \chi^2_{\mathrm{eff}} \sim 6$ with  $\ell_{\mathrm{max}}=2500$. Note that the better fit at $\ell=1000$ is also unconstrained by WMAP. 

 \subsection{Linear-spaced oscillations}
 
\begin{figure}[htbp] 
   \centering
   \includegraphics[width=3.4in]{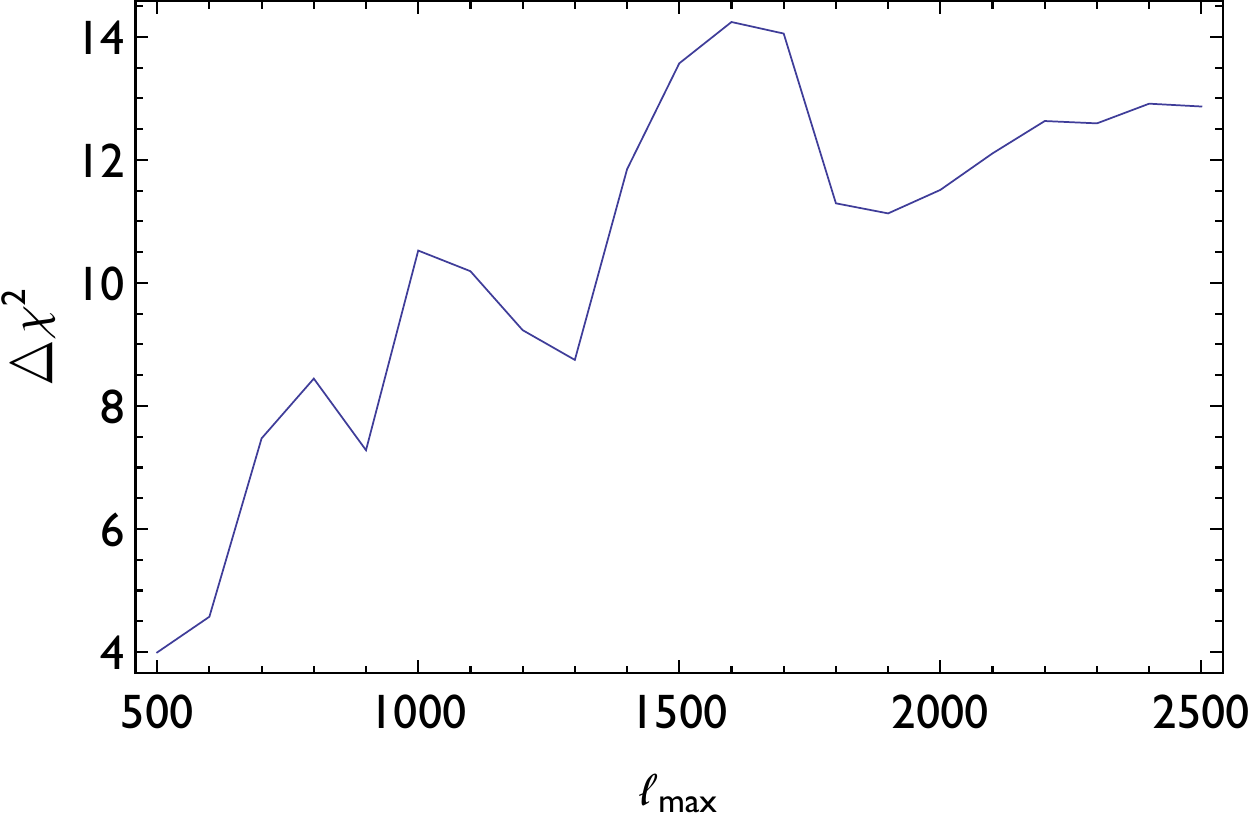} 
   \caption{This figure shows the improvement of the fit versus a reference fit for linear spaced oscillations. As expected, because of the similarities between the two features in Planck and WMAP, the improvement of fit is due to a feature at low multipoles. The feature at  $\ell=1800$ in the 217 GHz has the opposite effect; after $\ell =1800$ the improvement of the fit decreases. }
   \label{fig:chivslmaxlin}
\end{figure}

\begin{figure}[htbp] 
   \centering
   \includegraphics[width=3.4in]{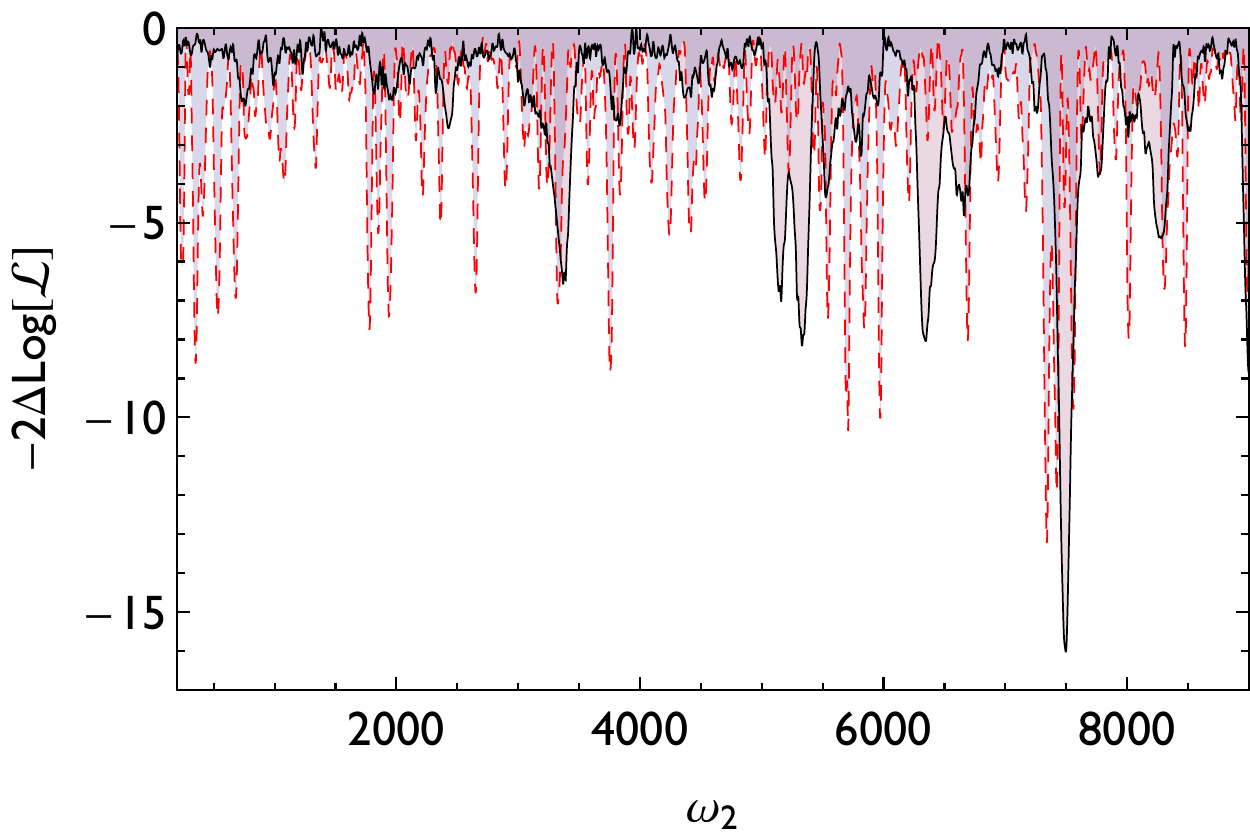} 
   \caption{Comparission between the improvement of fit for linear oscillations as a function of frequency between WMAP 9 (solid, black) and Planck (dashed, red).}
   \label{fig:wmapvsplanck3}
\end{figure}

Interestingly, in comparison with WMAP, the Planck data seems to contain many more  low frequency features as shown in Fig. \ref{fig:wmapvsplanck3}. As was the case for WMAP, the Planck data shows that higher frequencies can result in bigger the improvements of the fit. In WMAP we found were able to identify a single frequency that appeared to be favored over other frequencies. Despite the difference, a high oscillation feature persists in Planck data although the frequency has shifted slightly ($7500\rightarrow 7340$). Because of the similarities between the two data sets, this feature is most likely not due to a feature at small angular scales.  For this reason we again investigate the improvement of fit as a function of $\ell_{\mathrm{max}}$. This is shown in Fig.~\ref{fig:chivslmaxlin}. Interestingly, this fitting appears rather gradual, which would favor a true feature interpretation. Regarding the feature in the 217 GHz map, the improvement of the fit actually decreases after $\ell = 1800$. The plot shows that most of the improvement comes from  low multipoles, consistent with this feature appearing both in WMAP and in Planck at a similar frequency.

\begin{table*}
\centering
\begin{tabular}{| l | c | c | c | c | c | c | c | c | c | c | r |}
\hline\hline 
Parameter & $\omega_1/\omega_2$ & $\Omega_b h^2$ & $\Omega_c h^2$& $\tau$ & $H_0$ & $n_s $ &$\log 10^{10} A_s$ & $A_2/B_2$ & $\phi_1/\phi_2$ &$ A_{\mathrm{lens}} $ \tabularnewline
\hline 
Best-fit (log) & 13.2  &$0.022036$& 0.11661 & 0.083943 & 68.6 & 0.963 &3.19 & 0.022 & -0.48704 & 1.23\tabularnewline
Best-fit (lin) & 7340 &$0.021877$& 0.12003 & 0.079958 & 67 & 0.956 & 3.21& 0.179 & -0.448 & 1 (fixed) \tabularnewline
\hline
\end{tabular}
\caption{Best fit parameter values for log and linear spaced oscillations. }
\label{tab:bestfitlogplanck1}
\centering
\end{table*}

\section{Discussion and Conclusions} \label{conclusion}
In this second of two papers we have applied our recently developed code to search for resonant features in the Planck data. Our code recovers the results found by the Planck collaboration, but adds to these findings by significantly extending the frequency range of the search. In addition, our method finds larger improvements of fit for low frequencies because it varies all parameters to find the best fit, not just the amplitude and frequency of the oscillatory signal.

Our analysis has given us several important insights. First of all, the improvement at the low frequency of logarithmically spaced oscillations end are caused or at least enhanced by a varying lensing amplitude. For example, in comparison with the Planck paper result, we find that allowing the  lensing amplitude to vary shifts the best fit frequency to lower values, and improves the overall fit. For linear spaced oscillations we find the largest improvement at the highest frequencies, with a best fit frequency that is close to that  found in WMAP9. We showed that including this feature mostly improves the fit to the spectrum below $\ell \sim 1000$. 

Further comparison between our WMAP9 and Planck analyses, shows the improvement of fit for log spaced oscillations has flipped, i.e. while for WMAP9 the improvements were at high frequencies for Planck the best fit is at low frequencies, although a feature at high frequencies does appear in the Planck data. The fact that none of the found oscillations in Planck are present in WMAP9, suggested that most improvement is coming from high $\ell$, which we confirmed by computing the improvement as a function of $\ell_{\mathrm{max}}$. As the Planck team has noted, there is a feature present near $\ell = 1800$ in the $217\times217$ GHz spectrum that does not appear in the other frequencies. Our results suggest that the improvement at low frequencies is predominately due to this feature. 

\begin{figure}[htbp] 
   \centering
   \includegraphics[width=3.4in]{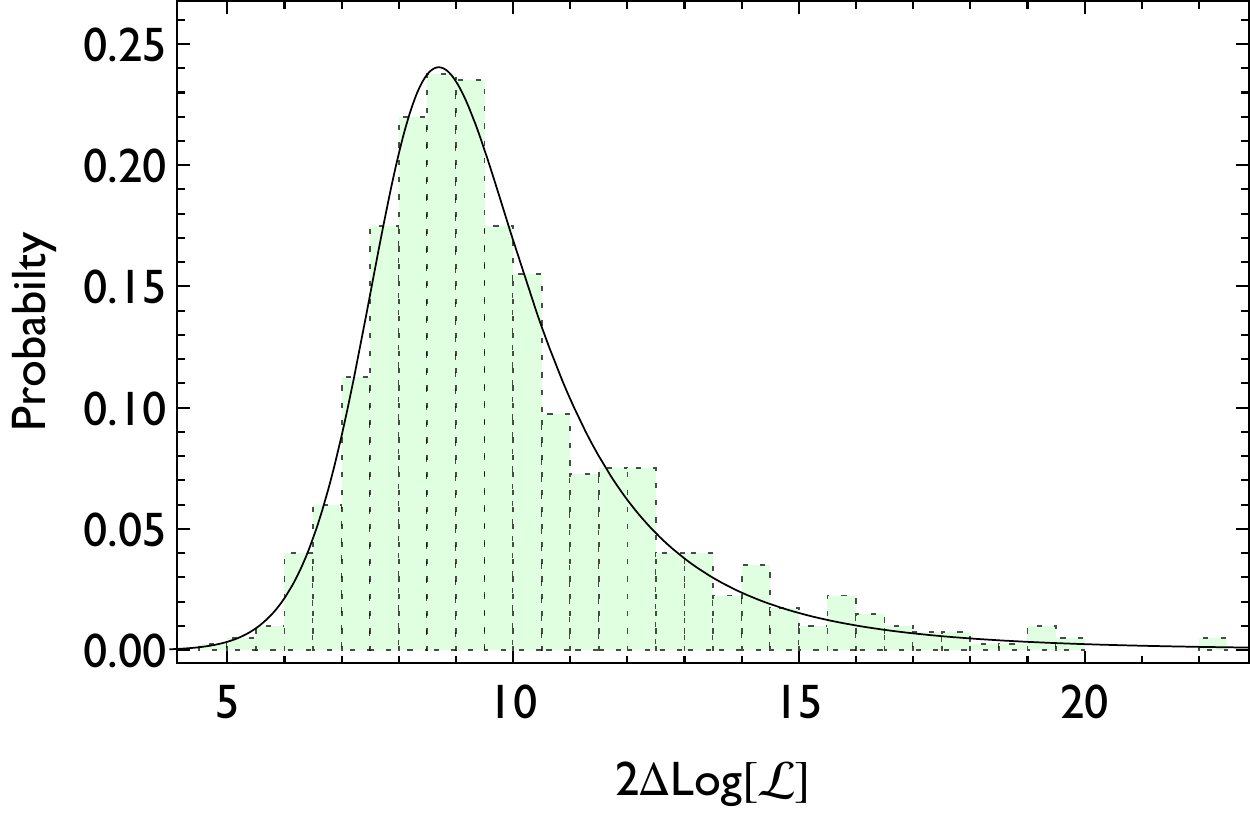} 
   \caption{Derived $\Delta \chi^2$ distribution for log spaced oscillations derived from simulating Gaussian Planck noise directly from the diagonal data covariance. The best fit were found using $-\pi \leq \phi_1 \leq \phi$ ($\Delta \phi_1 = \pi/3$),  $10 \leq \omega_1 \leq 250$ ($\Delta \omega_1 = 1$) and $0 \leq A_2^{\mathrm{eff}} \leq 0.06$ ($\Delta A_2^{\mathrm{eff}}= 0.005$) and $\ell \leq 32 \leq 2479$.}
   \label{fig:chidistribution}
\end{figure}

Second, in our companion paper we  tested our method on simulated data. The primary goal of these tests was to assess  the  reliability of our perturbative method. Here we use the same simulations to assess the significance of the fit improvements to determine  whether we have detected an oscillatory contribution to the primordial power spectrum. We found that for amplitudes as small as those that best fit the data, we expect an improvement of fit that exceeds what we find the data. We ran two full pipelines on maps that did not include a signal, for which we found improvements up to $\Delta \chi^2_{\mathrm{eff}} \sim 10$ \footnote{These simulations were performed fixing all $\Lambda$CDM parameters except for $\Omega_bh^2$.}. Furthermore we also ran a large number of (simplified) simulations in order to address the question: {\it What is the typical maximum improvement expected from fitting the noise ?} This analysis showed that the noise typically leads to $\Delta \chi^2_{\mathrm{eff}} \sim 10$, and has the potential to improve the fit $\Delta \chi^2_{\mathrm{eff}} \sim 25$. We found these improvements with $\ell_{\mathrm{max}} = 500$. We have improved on these simulations, by randomly generating Gaussian noise using the weighted error bars directly synthesized from the Planck data for multipoles ranging from $\ell=32$ to $\ell=2479$. We ran 1000 of these higher resolution featureless simulations and found that applying both log and linear spaced templates showed the measured (data) improvements are in the 90-94 percentile range. To be more precise, $\Delta \chi^2\simeq25$ is at the 3$\sigma$ level. These simulations are grid based, with  6 points in phase space, 12 points in the amplitude and 240 and 220 respectively for log and linear spaced oscillations in the frequency parameter. As a result, the derived distribution is conservative and not as accurate as a full simulation running an MCMC; with higher resolution we expect a distribution skewed towards higher improvements. Note that the resulting distribution is not a $\chi^2$ distribution, because this is a highly non-linear problem. E.g. assuming a cosmic variance limit experiment and a template that oscillates as $C_{\ell}\times g_{\ell}(A,\omega,\phi)$, we find 
\be
\Delta \chi^{2,i}(A,\omega,\phi) &=& \frac{f_s}{2}\sum_{\ell} (2\ell+1) \nonumber \\
&&\left[g_{\ell}^2-2^{3/2}f_s^{-1} g_{\ell}R_{\ell}^i(2\ell+1)^{-1/2}\right]\nonumber.
\ee
Here $f_s$ is the sky fraction and  $R_{\ell}^i$ is a Gaussian random variable with variance 1, independently drawn for each $\ell$ and Universe $i$. $g_{\ell}$ is oscillating and and contains 3 free parameters. To get the best-fit distribution for $\Delta \chi^2$ one would need to do an integral over the random variable, with best-fit $g_{\ell}$ which will now be a function of that same random variable. (see Fig.~\ref{fig:chidistribution}). This result  together with the full simulation using our code on mock data in the previous paper, suggests that  the improvements in fit found in our Planck analysis are consistent with expected statistical fluctuations for a realization from a featureless primordial model.

Third, both for linear and log spaced oscillations, improvements appear local in $\ell$ space. One might expect a real oscillation to lead to an improvement that would be more gradual as a function of the number of modes observed, though we recognize that this is not a rigorous argument:  the biggest improvements arise from  the modes with the highest signal to noise.

Lastly,  we have studied linear oscillations with the frequency at which the Planck team found a $3\sigma$ detection in the bispectrum \citep{2013arXiv1303.5084P}. We found that this frequency is very close to the Baryon Acoustic Oscillation (BAO) and we find no evidence in the power spectrum that there is such an oscillations (roughly corresponding to $\omega_2=220$). Our current understanding would suggest that varying the BAO parameters in the search for features in the bispectrum would probably reduce the significance (in addition to look elsewhere effects). 

For the future, we plan to implement Multinest as our sampler as  this will significantly speed up the code. As we go to higher frequencies we should include higher order terms to the derivative. In particular, for an accurate measurement of  $\Omega_b$, $\Omega_c$ and $H_0$ it may be necessary to include derivatives of these parameters in the expansion. 

When the Planck polarization data will be available, we should be able to improve
our search.  Oscillations should produce features in both temperature and polarization spectra (and cross-spectra).  Potentially, polarization measurements from ground-based experiments can probe out to $l \simeq 4000$ enabling even more sensitive searches for oscillatory features. However, searches based on ground-based data would be limited to lower oscillation frequencies since the power spectrum likelihood will have $\ell-\ell'$ correlations due to mode coupling effects.
We can get additional insight from three point measurements as models that predict oscillations in the power spectrum typically also predict oscillations in higher order correlation spectra (see e.g. Refs.~\citep{2007JCAP...06..023C,2011JCAP...01..017F,2009JCAP...05..018M,2010JCAP...02..001M,2011PhRvD..83d3520M}). While
there have been initial attempts to search for oscillations
in the Planck bispectrum measurements \citep{2013arXiv1303.5084P}, computational cost has limited these searches to low frequencies. Alternative approaches have been proposed to optimize this search \citep{2010PhRvD..82f3517M,2013arXiv1303.3499J}, but as of yet no attempt has been made to cover a large range of frequencies and phases. A first step would be to search the bispectrum measurements at frequencies suggested
by analyses of the CMB power spectrum.  Detecting features in both spectra would improve the statistical significance of the result---a promising direction
for future study.

\section*{Acknowledgments}
We would like to thank Fabian Schmidt, Renee Hlozek and Kendrick Smith for useful discussions. The authors would especially like to thank Raphael Flauger and Guido D'Amico for discussion and comments on the manuscript. P.D.M is supported by the Netherlands Organization for Scientific Research (NWO), through a Rubicon fellowship. P.D.M. and D.N.S. are in part funded by the John Templeton Foundation grant number 37426.

\bibliography{paper1_final}
  
\end{document}